\documentclass[aps,prb,reprint,groupedaddress]{revtex4-2}
\usepackage{amsmath}
\usepackage{bm} 
\usepackage{amssymb}
\usepackage{color}
\usepackage{multirow}
\usepackage{url}
\usepackage[colorlinks,bookmarks=false,citecolor=red,linkcolor=blue,urlcolor=blue]{hyperref}
\usepackage{graphicx}
\usepackage{mathrsfs,bbm}
\usepackage{wrapfig}
\usepackage{physics}
\usepackage{enumerate}
\usepackage[version=3]{mhchem}%
\usepackage{gensymb}
\usepackage{rotating}
\usepackage{multirow}

\def\k{\vec{k}}
\def\ahat{\hat{a}}
\def\bhat{\hat{b}}
\def\chat{\hat{c}}

\def\Hhat{\hat{H}}
\def\Hcal{\mathcal{H}}

\begin{document}
\title{Theoretical investigation of quantum oscillations of specific heat in Kondo insulators}
\author{Arnav Pushkar}
\author{Brijesh Kumar}
\email{bkumar@mail.jnu.ac.in}
\affiliation{School of Physical Sciences, Jawaharlal Nehru University, New Delhi 110067, India.}
\date{\today} 

\begin{abstract}
The electronic specific heat of Kondo insulators in magnetic field is studied for the half-filled Kondo lattice model on simple cubic lattice using a low-temperature theory in Kumar representation. The calculated specific heat is found to show quantum oscillations, which appear soon after the inversion transition and become prominent with decreasing Kondo coupling. Interestingly, it is noted that the field derivative of specific heat closely resembles the magnetic quantum oscillations, and exhibits more pronounced oscillations at finite temperatures than the magnetization itself. An empirical Lifshitz-Kosevich fit with two frequencies given by the theory describes these quantum oscillations reasonably well, where the frequencies correspond to the extremal areas on the surface of charge gap, a remnant of the Fermi surface in the insulating case. 
\end{abstract}

\maketitle

\section{Introduction}
Quantum oscillations in Kondo insulators has been a topic of much discussion in recent years. The magnetic quantum oscillations, normally considered a property of metals~\cite{shoenberg_1984}, observed in the Kondo insulators \ce{SmB6}~\cite{li2014two,tan2015unconventional,hartstein2018fermi} and \ce{YbB12}~\cite{Liu_2018,Xiang_2018,Ong_2018} present a very perplexing situation. Various theories have been put forward to resolve this~\cite{Knolle2015,Zhang2016,Erten2016,Pal2016,Peters2019,Sodemann2018,Lu2020,Tada2020,varma2020majoranas,Devakul2021}. The theory developed by Kumar and coworkers in Refs.~\cite{Ram2017,Ram2019,pushkar2023low} is particularly notable for its precise underpinning of this problem. According to this theory, the Kondo insulators come in two types distinguished by many-body inversion, and the quantum oscillations occur only in the inverted Kondo insulators.
Here the many-body inversion refers to a characteristic change with interaction in the dispersion of the charge quasiparticles such that the charge gap, which for strong Kondo couplings comes from the zone centre, 
shifts to a surface by inverting the dispersion around the zone centre upon decreasing the Kondo interaction below a critical strength. This theory finds magnetic quantum oscillations in the bulk of only such Kondo insulators which realize inverted quasiparticle dispersion, and not in the strong coupling Kondo insulators. It thus identifies the many-body inversion as a key determinant for the quantum oscillations to occur in Kondo insulators, and anticipates the surface of charge gap to act akin to a  Fermi surface for these quantum oscillations.

We further realized that the inversion is accompanied naturally by a dimensional reduction of the quasiparticle dispersion near the charge gap, with measurable signatures in quantities such as the density of states and specific heat. In going across the inversion transition from strong to weak Kondo couplings, the quasiparticle density of states near the charge gap, $\Delta_c$, changes from $(\epsilon-\Delta_c)^{1/2}$ (three dimensional) to $(\epsilon-\Delta_c)^{-1/2}$ (effectively one dimensional); correspondingly, the low-temperature behaviour of the specific heat changes from $T^{-1/2}\,e^{-\Delta_c/k_B T}$ for strong couplings to $T^{-3/2}\,e^{-\Delta_c/k_B T}$ for the inverted Kondo insulators~\cite{pushkar2023low}. In this paper, we extend this study further by investigating the specific heat as a probe of the quantum oscillations in Kondo insulators. Recent experiments on \ce{SmB6} have indeed measured the quantum oscillations of specific heat~\cite{labarre2022magnetoquantum}.

We calculate the specific heat, $C_v$, for the half-filled Kondo lattice model on simple cubic lattice as a function of the magnetic field, $B$. The key takeaways from these calculations are as follows: first, we find clear quantum oscillations of specific heat for the inverted Kondo insulators, much like the magnetic quantum oscillations; second, we realize that the field derivative of specific-heat, i.e. $\frac{1}{B}\frac{\partial C_v}{\partial B}$ vs. $\frac{1}{B}$, makes for a better presentation of the specific heat quantum oscillations, because it relates nicely to the magnetic quantum oscillations; third, the specific heat oscillations stay robust with increasing temperature whereas the magnetic oscillations die off quickly; fourth, the Lifshitz-Kosevich type behaviour with respect to thermal as well as quantum fluctuations is exhibited by the specific heat oscillations, but not for very low temperatures or high Kondo couplings; and fifth, the idea that the frequencies of quantum oscillations come from the surface of charge gap is supported by the specific heat oscillations. All these points are relevant to experiments. 

Section~\ref{sec:model} of this paper describes the model and the method of calculation. In Sec.~\ref{sec:QO-cv}, we present and analyze the calculated data for the quantum oscillations of specific heat, and compare it with magnetic oscillations obtained form the earlier calculations in Ref.~\cite{pushkar2023low}. We perform an empirical two-frequency Lifshitz-Kosevich fit in Sec.~\ref{sec:freq}, which describes the calculated quantum oscillations reasonably well and provides a support for the idea that their frequencies correspond to the extremal areas of the orbits on the surface of charge gap. Section~\ref{sec:sum} concludes the paper with a summary.

\section{\label{sec:model} Model and Method}
To investigate the quantum oscillations of specific heat in Kondo insulators, 
we study the half-filled Kondo lattice model (KLM) on simple cubic lattice. It is a basic model of Kondo insulators given by the Hamiltonian, $\mathcal{H}$, written below.  
\begin{equation}
\Hcal = -t\sum_{\vec{r},\vec{\delta}}\sum_{s=\uparrow,\downarrow}\chat_{\vec{r},s}^\dag \chat_{\vec{r}+\vec{\delta},s}+\frac{J}{2}\sum_{\vec{r}} \vec{S}_{\vec{r}} \cdot \vec{\tau}_{\vec{r}}
\label{eq:H}
\end{equation} 
It describes localized spin-1/2 moments (Pauli operators, $\vec{\tau}_{\vec{r}}$) coupled antiferromagnetically ($J>0$) with itinerant electrons given by the creation (annihilation) operators, $\chat^\dag_{\vec{r},s} (\chat^{ }_{\vec{r},s})$, and the spin operators, $\vec{S}_{\vec{r}}$. The electrons are given to hop with a nearest-neighbour amplitude, $t$, on a simple cubic lattice of $L$ sites given by the position vectors, $\vec{r}$. The six nearest-neighbours of a site are given by $\vec{\delta}=\pm a\hat{x}, \pm a\hat{y}, \pm a\hat{z}$, where $a$ is the lattice constant. 
 
A theory of Kondo insulators was developed in Ref.~\cite{Ram2017} 
by studying the charge and spin dynamics of the half-filled KLM self-consistently in Kumar representation~\cite{Kumar2008}. This theory discovered many-body inversion exhibited by the quasiparticle dispersion with decreasing Kondo coupling (i.e. increasing $t/J$), and found magnetic quantum oscillations in the Kondo insulating ground state for moderate to weaker Kondo couplings. This theory was further extended to finite temperatures in Ref.~\cite{pushkar2023low}, where the inversion was found to occur also at finite temperatures and the magnetic quantum oscillations were found to behave in a Lifshitz-Kosevich like manner not only with temperature but also with quantum fluctuations (characterized by an RKKY like parameter $J^2/t$). There it was also noted that the inversion comes with an effective dimensional reduction of the quasiparticle dispersion near the charge gap, with measurable consequences for the density of states and specific heat as mentioned in the introduction. 
Below we give an outline of this theory. 

In Kumar representation~\cite{Kumar2008}, we can write the electron operators as: $\chat^\dag_{\vec{r}\uparrow} = \hat{\phi}_{a,\vec{r}} \, \sigma^+_{\vec{r}}$, $\chat^\dag_{\vec{r} \downarrow} = \frac{1}{2}(i\psi_{a,\vec{r}}-\phi_{a,\vec{r}}\,\sigma^z_{\vec{r}})$ on $A$-sublattice, and $\chat^\dag_{\vec{r},\uparrow} = i\hat{\psi}_{b,\vec{r}}\,\sigma^+_{\vec{r}}$, $\chat^\dag_{\vec{r},\downarrow} = \frac{1}{2}(\phi_{b,\vec{r}}-i\psi_{b,\vec{r}}\,\sigma^z_{\vec{r}})$ on $B$ sublattice. Here $\sigma^\pm_{\vec{r}}$ and $\sigma^z_{\vec{r}}$ denote the Pauli operators, whereas $\phi_{a,\vec{r}} = \ahat^{ }_{\vec{r}}+\ahat^\dag_{\vec{r}}$ and $\psi_{a,\vec{r}}=i(\ahat^{ }_{\vec{r}} - \ahat^\dag_{\vec{r}})$ are the Majorana operators corresponding to the spinless fermions, $\ahat^{ }_{\vec{r}}$, on $A$-sublattice, and likewise for the spinless fermions, $\bhat^{ }_{\vec{r}}$, on $B$-sublattice. The electrons are thus described canonically in terms of spinless fermions and Pauli operators.  We write Eq.~\eqref{eq:H} in this representation and then decouple it as: $\Hcal=\Hcal_c+\Hcal_s+e_0L$, into a spinless fermion part, $\Hcal_c$, describing the charge dynamics self-consistently with another part, $\Hcal_s$, describing the spin dynamics; $e_0L$ is the decoupling constant. We skip the technical details of this theory to avoid repetition and urge the interested readers to look at Refs.~\cite{Ram2017,pushkar2023low}. The theory not only describes the low-temperature insulating behaviour and the Kondo-singlet to antiferromagnetic transition, but it also reveals the inversion of charge quasiparticle dispersion with interaction. It finds that, as $t$ (in units of $J$; at a fixed temperature, $T$) increases beyond the `inversion' point $t_i$, the charge gap shifts from the zone centre to a surface around the zone centre, while the spins continue to form Kondo singlet. Upon increasing $t$ further, beyond a critical $t_c$, the Kondo singlet gives way to antiferromagnetic (AFM) order. Thus, according to this theory, the strong coupling Kondo singlet phase (KS) for $t<t_i$ is distinguished from the intermediate Kondo singlet phase (iKS) for $t_i < t < t_c$, and the AFM phase for $t>t_c$, by the inversion; refer to the phase diagram presented in Ref.~\cite{pushkar2023low}. Moreover, as the Kondo coupling gets weaker, the charge-gap surface tends to approach the non-interacting Fermi surface. 

To investigate the quantum oscillations, we introduce the magnetic field, $\vec{B}= B\hat{z}$, along $z$-direction in Eq.~\eqref{eq:H} through Peierls phase, $e^{i\frac{e}{\hbar} \int_{\vec{r}}^{\vec{r}+\vec{\delta}} \vec{A}\cdot \vec{dl}}$, in the hopping, with $\vec{A}=-yB\hat{x}$ as the vector potential. By the same steps as described above, we obtain the following effective model of charge dynamics for the half-filled KLM in magnetic field; see Ref.~\cite{Ram2017} for the derivation. 
\begin{align}
\mathcal{H}_{c,\alpha} & = \frac{J\rho_0}{4}\left(\sum_{\vec{r}\in A}\ahat^\dag_{\vec{r}}\,\ahat^{ }_{\vec{r}}+\sum_{\vec{r}\in B}\bhat^\dag_{\vec{r}}\,\bhat^{ }_{\vec{r}}\right) -\frac{it}{2}\sum_{\vec{r}\in A}\sum_{\vec{\delta}} \nonumber \\ 
& \cos(2\pi\alpha \, r_y \, \hat{x}\cdot\hat{\delta}) \left[\psi^{ }_{a,\vec{r}}\,\phi_{b,\vec{r}+\vec{\delta}}+\rho_1\psi_{b,\vec{r}+\vec{\delta}} \, \phi^{ }_{a,\vec{r}}\right]
\label{eq:Hcalpha}
\end{align}
Here $\rho_0$ is a measure of the Kondo singlet, $\rho_1$ is a measure of the antiferromagnetic correlation between nearest-neighbours, $\alpha=Ba^2/(e/h)$ is the magnetic flux in units of $e/h$, $r_y$ is the $y$ component of the position vector $\vec{r}$ of the sites on the $A$ sublattice, and $\vec{\delta}$ is summed over the nearest neighbours. In calculations, $\alpha$ is taken as $p/q$ with $p=1,2\dots (q+1)/2$ for a prime number $q$; we take $q=401$ in the present calculations. We diagonalize $\Hcal_{c,\alpha}$ numerically for different values of $\alpha$ for several different temperatures and hopping by putting in Eq.~\eqref{eq:Hcalpha} the values of $\rho_0$ and $\rho_1$ obtained from the zero-field self-consistent equations in Ref.~\cite{pushkar2023low}. From this, the magnetization is computed as the field derivative of the free energy. This is how the magnetic quantum oscillations were calculated for the half-filled KLM in Refs.~\cite{Ram2017, pushkar2023low}. Below we compute the specific heat as a function of magnetic field using Eq.~\eqref{eq:Hcalpha} to see if it exhibits oscillations and how it compares with magnetic quantum oscillations.

\section{Quantum Oscillations of Specific Heat}\label{sec:QO-cv}
We calculate the specific heat from the internal energy per site, $u = \langle \Hcal \rangle / L$ by taking its temperature derivative, $C_v=\partial{u}/\partial{T}$. With $\Hcal$ decoupled into $\Hcal_c$ and $\Hcal_s$, as described earlier, it turns out that $u = \langle \Hcal_c \rangle/L$. In the presence of magnetic field, i.e. for $\alpha\neq 0$, we diagonlize Eq.~\eqref{eq:Hcalpha} numerically and compute the field dependent internal energy as $u_\alpha=\langle \Hcal_{c,\alpha}\rangle/L$, the formal expression for which can be written as:
\begin{equation}
	{u}_\alpha=\frac{J\rho_0}{8}-\frac{1}{2L}\sum_{\vec{k}}\sum_{\nu=1}^{2q}\epsilon_{\vec{k},\nu}\tanh \left(\frac{\beta\epsilon_{\vec{k},\nu}}{2}\right)
	\label{eq:Ucalpha}
\end{equation}
where $\epsilon_{\vec{k},\nu}$ are the quasiparticle energies and $\vec{k} = (k_1,k_2,k_3)$ for $k_{1(3)} \in [-\pi,\pi]$ and $k_2 \in [-\frac{\pi}{q},\frac{\pi}{q}]$. Refer to the Appendix for details on this. In our calculations, $J=1$. We compute this internal energy as a function of $\alpha$ for many different temperatures and hopping amplitudes. From this we obtain $C_v=\partial{u_\alpha}/\partial{T}$ as a function of $\alpha$ over a range of $T$ and $t$. The findings from these calculations are presented below. 

\begin{figure}[t]
	\centering
	\includegraphics[width=\columnwidth]{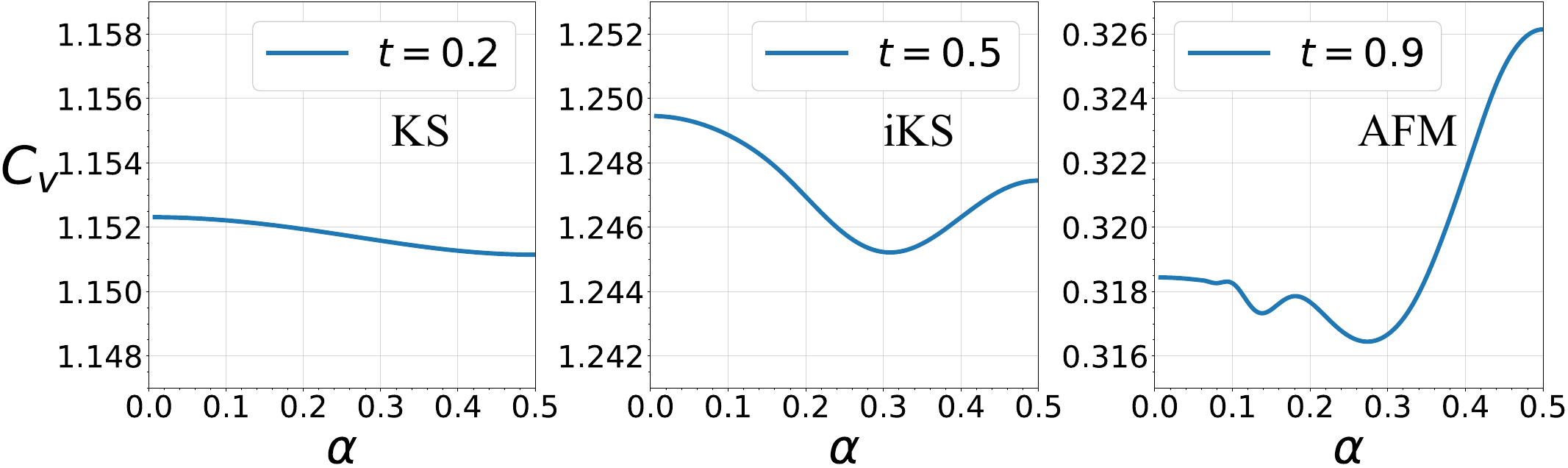}
	\caption{Specific heat, $C_v$, vs. magnetic field, $\alpha$, calculated for the half-filled Kondo lattice model on simple cubic lattice at temperature, $T=0.2$, for three representative values of the hopping: $t=0.2$ in the strong coupling Kondo singlet (KS) phase, $t=0.5$ in the inverted Kondo singlet (iKS) phase, and $t=0.9$ in the antiferromagnetic (AFM) phase. Note that, for $T=0.2$, the inversion occurs at $t_i=0.3$.}
	\label{fig:CvOsc}
\end{figure}

Figure~\ref{fig:CvOsc} presents the specific heat data, as obtained from our calculations at a fixed temperature, for three different values of $t$ representing the KS, iKS and AFM phases. For the strong coupling KS case, the $\alpha$ dependence of $C_v$ is weak and structureless with no signs of oscillations. In the iKS phase for moderate couplings, $C_v$ varies non-monotonically with $\alpha$, showing some signs of an oscillatory behaviour. The oscillations of specific heat become stronger by increasing $t$ further; deep inside the AFM phase, they are very prominent. Hence, the specific heat exhibits quantum oscillations only in the inverted (iKS and AFM) Kondo insulating phases. This is consistent with the magnetic quantum oscillations calculated earlier by the same theory in Refs.~\cite{Ram2017,pushkar2023low}.

\begin{figure}[htbp]
	\centering
	\includegraphics[width=\columnwidth]{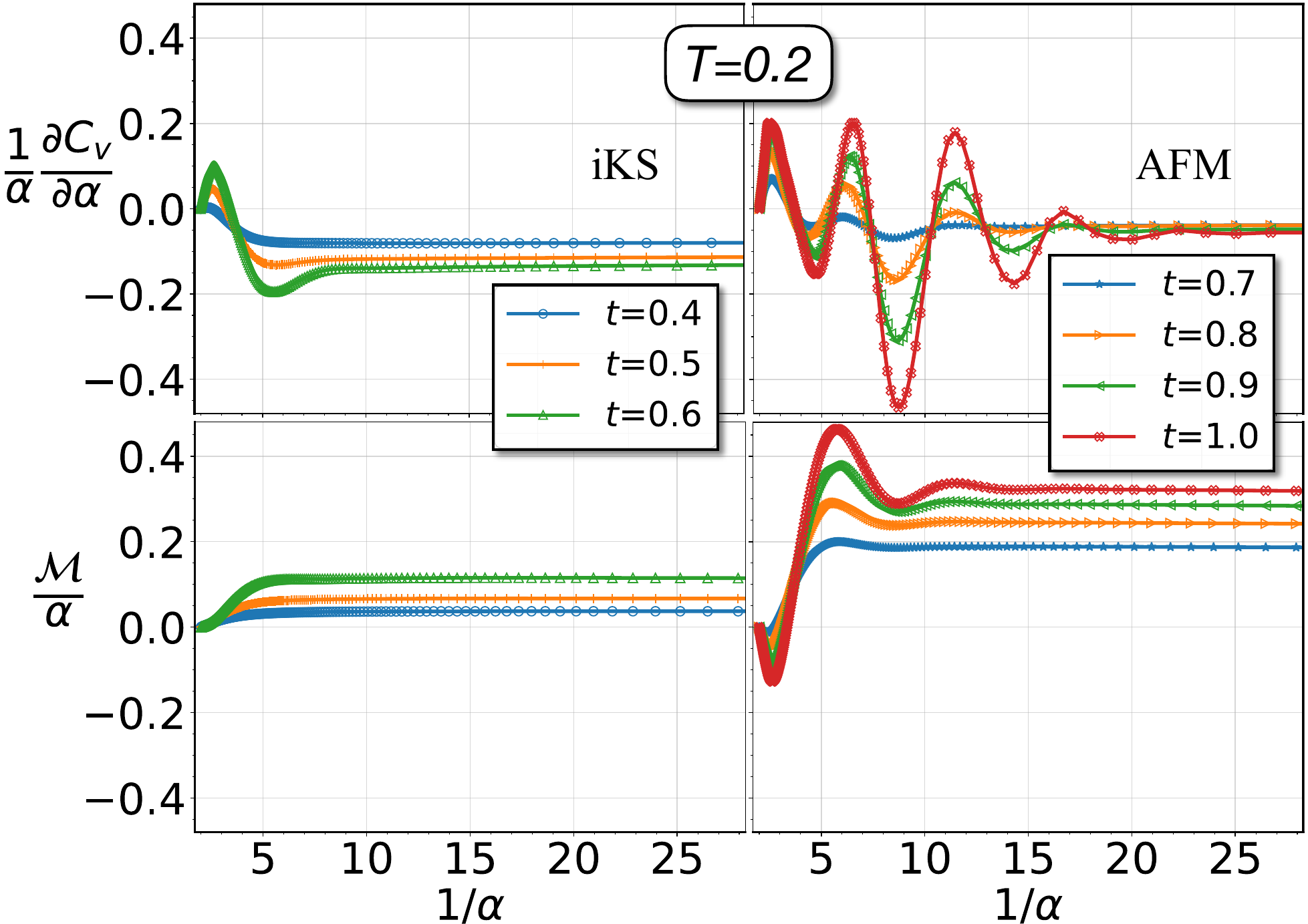}
	\caption{Magnetic field dependence of specific heat represented as $\frac{1}{\alpha}\frac{\partial C_v}{\partial \alpha}$ vs. $\frac{1}{\alpha}$ (upper two plots) for comparison with magnetization, $\frac{\mathcal{M}}{\alpha}$ vs. $\frac{1}{\alpha}$ (lower two plots), at $T=0.2$ for different values of $t$ in the iKS and AFM insulating phases.}
	\label{fig:OscComp}
\end{figure}

We realize that an interesting way of presenting the specific heat data, that can be compared directly with the magnetic quantum oscillations, is to plot $\frac{1}{\alpha} \frac{\partial C_v}{\partial \alpha}$ vs. $\frac{1}{\alpha}$. This is motivated by the following equation that relates specific heat to magnetization, $\mathcal{M}$.
\begin{align}
\frac{1}{\alpha}\frac{\partial C_v}{{\partial\alpha}} &= \frac{T}{\alpha}\frac{\partial^2 \mathcal{M}}{\partial T^2}
\label{eq:Cv_M}
\end{align} 
Figure~\ref{fig:OscComp} presents the specific heat data as $\frac{1}{\alpha} \frac{\partial C_v}{\partial \alpha}$ vs. $\frac{1}{\alpha}$, together with the magnetization data as $\frac{\mathcal{M}}{\alpha}$ vs. $\frac{1}{\alpha}$ from the same calculation; their mutual resemblance is remarkable. Notably, the quantum oscillations of specific presented in this manner look more prominent than those of magnetization. For instance, in the iKS phase, where the magnetic quantum oscillations are not quite formed, the specific heat oscillations are reasonably clear. We see the hint of oscillation in specific heat already at $t=0.4$, i.e. soon after the inversion as expected, and for $t=0.5$ and $0.6$, the oscillations are clearly present in specific heat but not in magnetization. Hence, through  $\frac{1}{\alpha}\frac{\partial C_v}{{\partial\alpha}}$, we get to see the quantum oscillations even better. 

\begin{figure*}[htbp]
	\centering
	\includegraphics[width=\textwidth]{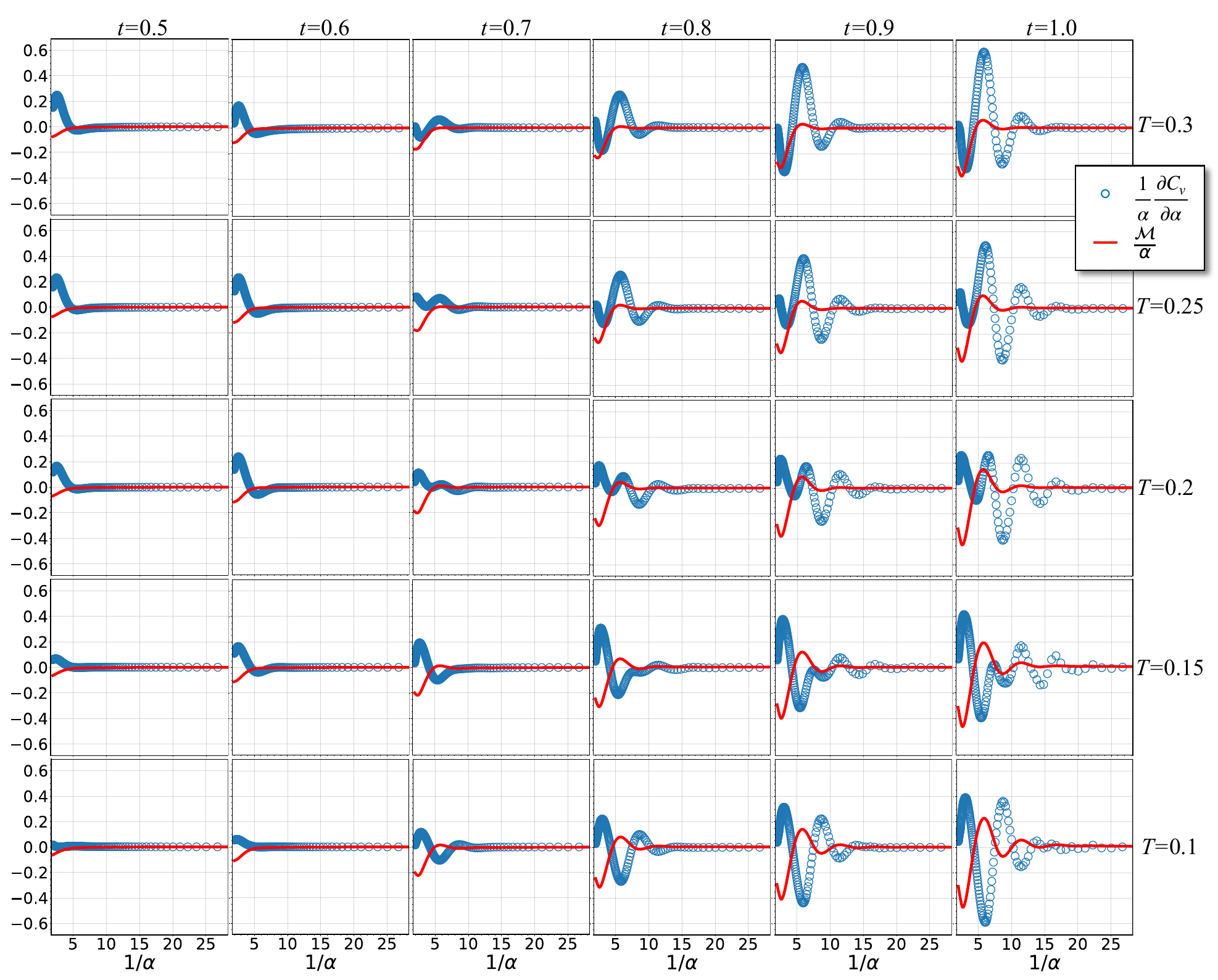}
	\caption{Quantum oscillations of specific heat (blue circles). Also shown alongside are the magnetic quantum oscillations (red lines). The specific heat oscillations are clearly more prominent over a wider range of $t$ and $T$ (in units of $J$). What is presented here are the oscillatory parts of $\frac{1}{\alpha}\frac{\partial C_v}{\partial \alpha}$ and $\frac{\mathcal{M}}{\alpha}$ after subtracting off the non-oscillatory background.}
	\label{fig:cv-M-compare-grid}
\end{figure*}

Note that, in Fig.~\ref{fig:OscComp}, the oscillations of $\frac{1}{\alpha}\frac{\partial C_v}{{\partial\alpha}}$ and $\frac{\mathcal{M}}{\alpha}$ tend to die out with increasing $\frac{1}{\alpha}$. To focus on the oscillatory part of these quantities, we subtract the  non-oscillatory background from the data. The quantum oscillations of specific heat calculated over a wide range of hopping and temperature are thus presented in Fig.~\ref{fig:cv-M-compare-grid}; the magnetic oscillations from the same calculation are also shown there for comparison. Similar to the magnetic oscillations, the specific heat oscillations too grow stronger as $t$ increases; however, the specific heat starts exhibiting oscillations at comparatively smaller values of hopping. Notably, the specific heat oscillations show up prominently at increased temperatures where the magnetic oscillations weaken considerably; this is because $C_v$ is enhanced by increasing $T$. It suggests that the specific heat could well be a better probe of quantum oscillations at finite temperatures. 

\begin{figure}[htbp]
	\centering
	\includegraphics[width=\columnwidth]{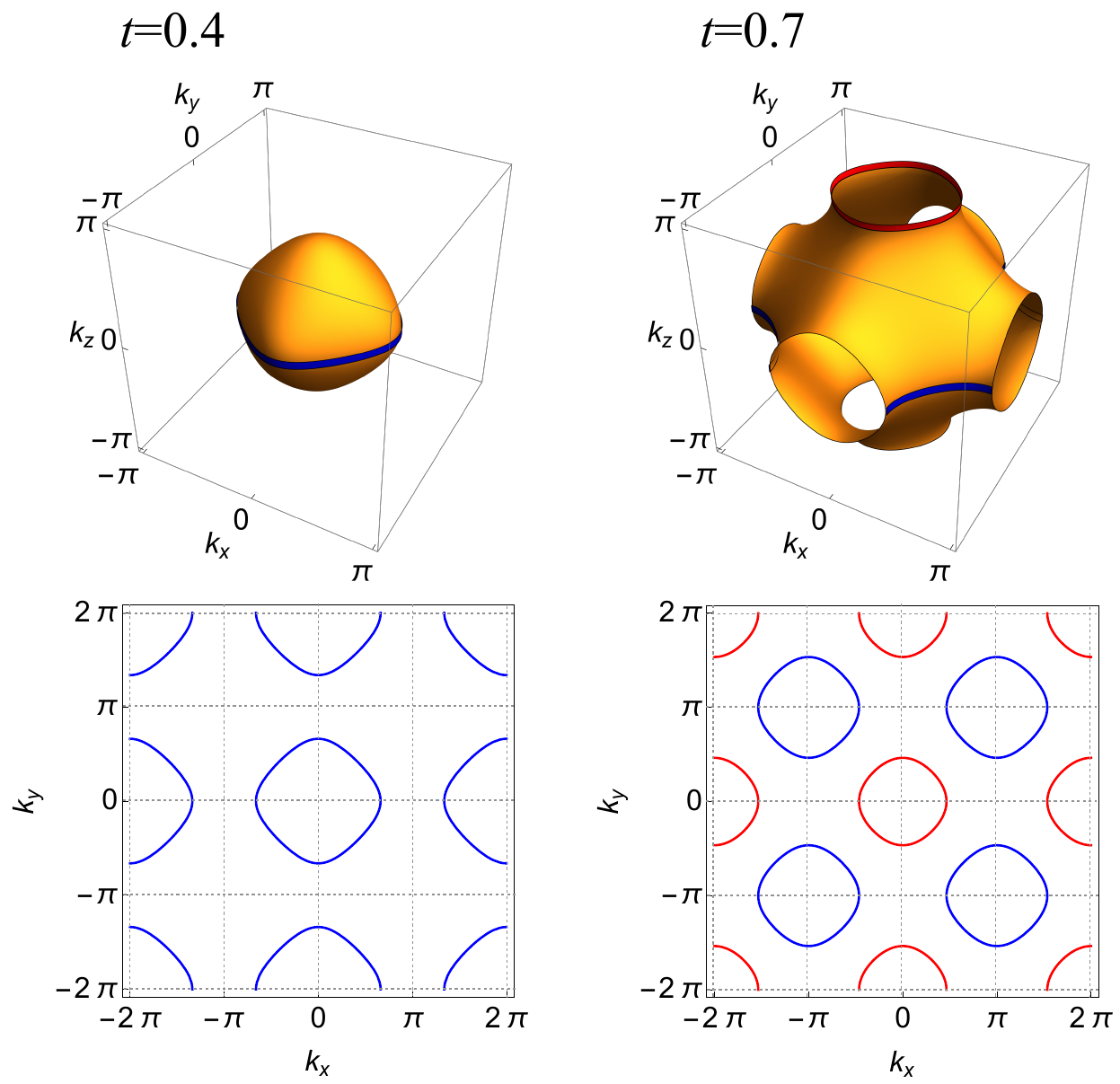}
	\caption{The surface of charge gap. For $t$'s closer to $t_i$ (the inversion point), it is closed around the zone centre as shown for $t=0.4$. Farther away from $t_i$, it changes into an open surface as for $t=0.7$. Closed orbits (with extremal areas in the planes $\perp$ $k_z$) on the charge-gap surface are shown in the respective two-dimensional plots. For $t=0.4$, there is only one such orbit on $k_z=0$ plane, and for $t=0.7$, there are two such orbits lying on $k_z=0$ (blue) and $k_z=\pi$ (red) planes.} 
	\label{fig:GapSurf}
\end{figure}

\section{\label{sec:freq} Oscillation Frequencies and the surface of Charge Gap}
Inversion is a key determinant for quantum oscillations to occur in the insulating bulk. As described earlier, in the two phases (iKS and AFM) with inverted quasiparticle dispersion, the charge gap comes from the surface  given by the following equation in the Brillouin zone~\cite{Ram2017}.
\begin{equation}
	\gamma_{\vec{k}}=\frac{J|\rho_0|(1+|\rho_1|)}{4t(1-|\rho_1|)\sqrt{|\rho_1|}}
	\label{eq:Gap_surface}
\end{equation}
Here $\gamma_{\vec{k}} = 2(\cos{k_x a} + \cos{k_y a} + \cos{k_z a})$. In the weak coupling limit, it tends to $\gamma_{\vec{k}} = 0$, the Fermi surface of the non-interacting tight binding model at half filling. Figure~\ref{fig:GapSurf} shows how it typically looks. For $t$'s near $t_i$, it is a closed surface around the zone centre, with one extremal orbit on $k_z=0$ plane. Sufficiently away from $t_i$, it changes into an open surface, with two extremal orbits on $k_z=0$ and $k_z=\pi$ planes. Let the area enclosed by the orbit on $k_z=0$ plane be denoted as $A_0$, and that on $k_z=\pi$ plane be denoted as $A_\pi$; see Table~\ref{tab:Area} for their values from our theory. Since the magnetic field in our calculations is taken along $z$-direction, we assume an Onsager like relation for these areas to identify possible frequencies of quantum oscillations. The areas listed in Table~\ref{tab:Area} in units of $(\frac{2\pi}{a})^2$ are same as the frequencies in units of $\frac{h}{ea^2}$. Below we analyze the calculated quantum oscillations using these frequencies. 

In Refs.~\cite{Ram2017,pushkar2023low}, by doing Fourier analysis, we estimated the frequency of magnetic quantum oscillations to be around 0.185, which is the area of the closed extremal orbit on $\gamma_{\vec{k}}=0$ surface, i.e. the surface of charge gap in the weak coupling limit. We note that 0.185 also happens to be the average of $A_0$ and $A_\pi$ for $t\gtrsim 0.7$. Since the quantum oscillations of specific heat calculated in the present work are more prominent than those of magnetization over a wider range of $t$ and $T$, here we have a scope to check if these oscillations are better described by the two frequencies, $A_0$ and $A_\pi$, coming from the surface of charge gap. To this end, we fit the specific heat oscillations empirically to the Lifshitz-Kosevich (LK) formula with two frequencies.  The LK formula~\cite{LK,shoenberg_1984} for the thermal behaviour of magnetic quantum oscillations can be written as:
	\begin{align}
		\mathcal{M}=\frac{T}{\sqrt{\alpha}}\sum_{\nu,n} c_{\nu,n} \frac{(-1)^{n+1}}{\sqrt{n}} \frac{\sin\left(\frac{2\pi\,n\,A_\nu}{\alpha}+\frac{\pi}{4}\right)}{\sinh\left(\frac{n\,b_{\nu}T}{\alpha}\right)}
		\label{eq:LK-M}
	\end{align}
where $A_\nu$ for $\nu=0,\pi$ are the frequencies (areas) given in Table~\ref{tab:Area}, and $c_{\nu,n}$ and $b_\nu$ are the fitting parameters. Integer $n$ takes the values $1,2,3,\dots,\infty$ for each $\nu$, but we keep only the leading few terms (minimum two and maximum four) for the purpose of fitting. 
By applying Eq.~\eqref{eq:Cv_M} to this LK formula for magnetization, we get
\begin{align}
	\frac{\partial C_v}{\partial\alpha}=&\frac{T}{\sqrt{\alpha}}\sum_{\nu,n} c_{\nu,n} \frac{(-1)^{n+1}}{\sqrt{n}}\frac{\partial^2}{\partial T^2}\left[\frac{T}{\sinh\left(\frac{n\,b_{\nu}T}{\alpha}\right)}\right]\nonumber \\
	&  \times\sin\left(\frac{2\pi \,n \, A_\nu}{\alpha}+\frac{\pi}{4}\right)
	\label{eq:LK-cv}
\end{align}
for the specific heat. In this expression, the thermal amplitude, $\frac{\partial^2}{\partial T^2} \left[\frac{T}{\sinh(n b_\nu T/\alpha)}\right]$, can change sign with $T$; this is unlike the amplitude, $\frac{T}{\sinh(n b_\nu T/\alpha)}$, for magnetization. This feature can be seen in Fig.~\ref{fig:cv-M-compare-grid}. By carefully looking there at the specific heat oscillations in any column, say for $t=0.9$, we see the peaks turn into troughs (and vice versa) with increasing $T$. It is an indication of the LK type behaviour present in the quantum oscillations of specific heat. Let us see how well the specific heat data in Fig.~\ref{fig:cv-M-compare-grid} is described by Eq.~\eqref{eq:LK-cv}, and to what extent.   

\begin{table}[htbp]
	\begin{tabular}{l||c|c|c|c|c|}
		$t \diagdown T$ & {\hspace{5mm} 0.1} & 0.15 & 0.2 & 0.25 & 0.3  \\ \hline \hline 
		0.4 & $\begin{array}{l l} A_0~ & 0.223 \\ A_\pi~ & 0\end{array}$ & $\begin{array}{l} 0.247 \\ 0\end{array}$ & $\begin{array}{l} 0.304  \\ 0\end{array}$ & $\begin{array}{l}0.39 \\ 0\end{array}$ & $\begin{array}{l}0.476 \\ 0.006\end{array}$ \\\hline
		0.5 & $\begin{array}{l l} A_0~ & 0.476 \\ A_\pi~ & 0.006\end{array}$  & $\begin{array}{l}0.442  \\ 0.017\end{array}$  & $\begin{array}{l}0.391 \\ 0.039\end{array}$  & $\begin{array}{l} 0.348 \\ 0.061\end{array}$ & $\begin{array}{l} 0.315 \\ 0.081\end{array}$ \\ \hline
		0.6 & $\begin{array}{l l} A_0~ & 0.29 \\ A_\pi~ & 0.097\end{array}$ & $\begin{array}{l} 0.277 \\ 0.107\end{array}$ & $\begin{array}{l}0.264 \\ 0.116\end{array}$ & $\begin{array}{l} 0.257 \\ 0.122\end{array}$ & $\begin{array}{l} 0.253 \\ 0.124\end{array}$ \\ \hline
		0.7 & $\begin{array}{l l} A_0 & 0.208 \\ A_\pi~ & 0.163\end{array}$ & $\begin{array}{l} 0.208 \\ 0.163\end{array}$ & $\begin{array}{l} 0.21\\ 0.161\end{array}$ & $\begin{array}{l}0.213 \\ 0.158\end{array}$ & $\begin{array}{l} 0.218 \\ 0.154\end{array}$ \\ \hline
		0.8 & $\begin{array}{l l} A_0~ & 0.198 \\ A_\pi~ & 0.172\end{array}$ & $\begin{array}{l} 0.198 \\ 0.172\end{array}$ & $\begin{array}{l} 0.198 \\ 0.172\end{array}$ & $\begin{array}{l} 0.2 \\ 0.17\end{array}$ & $\begin{array}{l} 0.203 \\ 0.167\end{array}$ \\ \hline
		0.9 & $\begin{array}{ll} A_0~ & 0.196 \\ A_\pi~ & 0.174\end{array}$ & $\begin{array}{l} 0.195 \\ 0.175 \end{array}$ & $\begin{array}{l} 0.195 \\ 0.175\end{array}$ & $\begin{array}{l} 0.196 \\ 0.174\end{array}$ & $\begin{array}{l}0.197 \\ 0.172\end{array}$ \\ \hline
		1.0 & $\begin{array}{ll} A_0~ & 0.195 \\ A_\pi~ & 0.175\end{array}$ & $\begin{array}{l} 0.194 \\ 0.176\end{array}$ & $\begin{array}{l} 0.194 \\ 0.176\end{array}$ & $\begin{array}{l}0.194 \\ 0.176\end{array}$ & $\begin{array}{l} 0.195 \\ 0.175\end{array}$ \\ \hline
	\end{tabular}
	\caption{Extremal areas of the contours on the surface of charge gap, Eq.~\eqref{eq:Gap_surface}, at different temperatures for different values of hopping. Here $A_0$ and $A_\pi$ denote the areas on $k_z=0$ and $k_z=\pi$ plane, respectively, and their values are given in units of $(2\pi/a)^2$ which are same as the frequencies of quantum oscillations in units of $h/ea^2$.}
	\label{tab:Area}
\end{table}

\begin{figure*}[htbp]
	\centering
	\includegraphics[width=\textwidth]{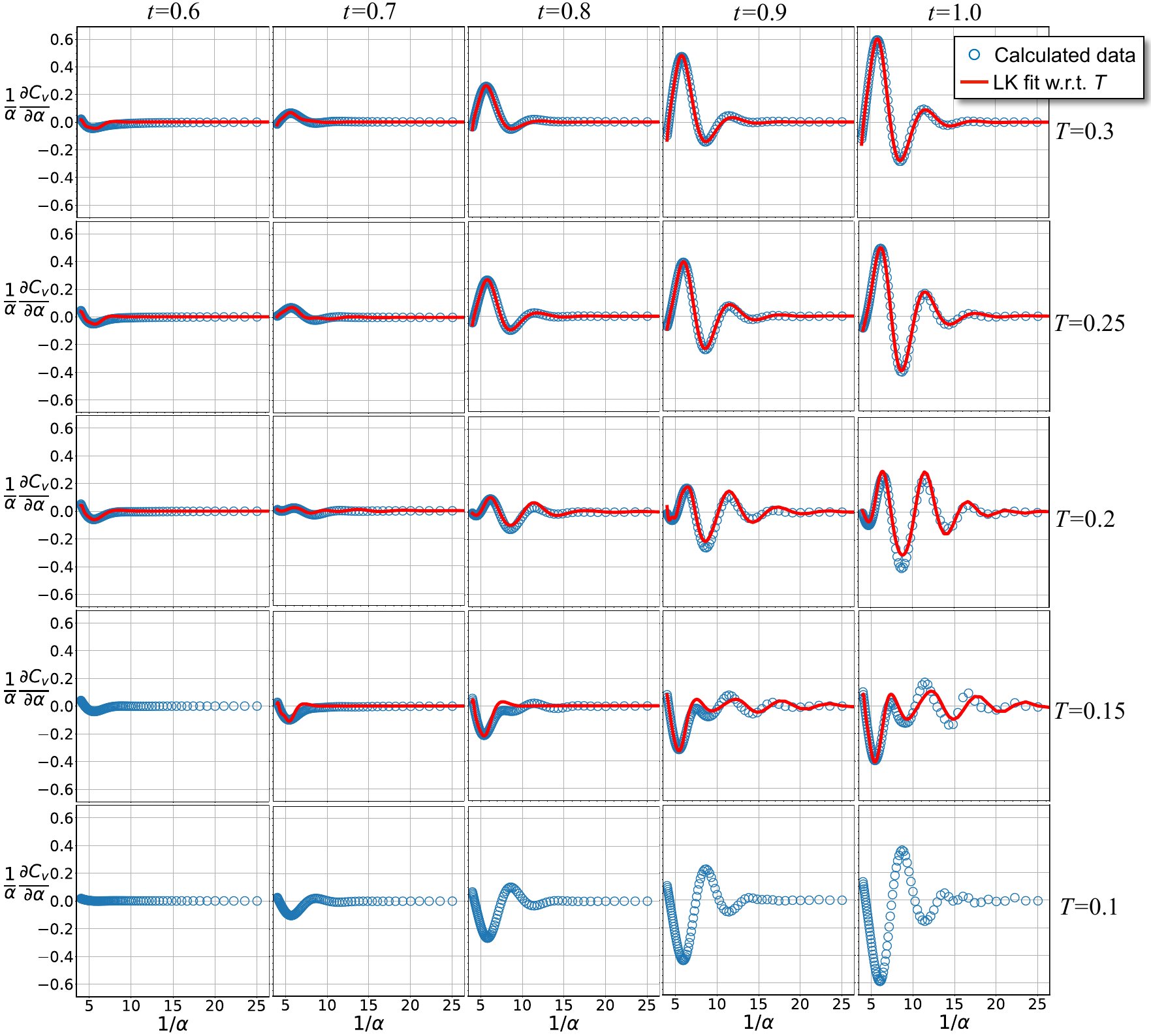}
	\caption{Empirical fitting (red lines) of the specific heat oscillations (blue circles) to the Lifshitz-Kosevich formula for  thermal behaviour, Eq.~\eqref{eq:LK-cv}, with two frequencies ($A_0$ and $A_\pi$ listed in Table~\ref{tab:Area}) corresponding to the extremal areas on the surface of charge gap. Here (and in Figs.~\ref{fig:LK_T_mag} to~\ref{fig:LK_tinv_mag}) the plots without the red lines are those where the fitting did not work.}
	\label{fig:LK_T_Cv}
\end{figure*}

In Fig.~\ref{fig:LK_T_Cv}, we present the fitting of the specific heat oscillations to the above LK formula with frequencies taken from Table~\ref{tab:Area}, for several different values of hopping ($t=0.6,0.7,0.8,0.9,1.0$) and temperature ($T=0.1,0.15,0.2,0.25,0.3$). Let us look closely at each row of plots at a given temperature. Starting from the top row, we see that the fitting is visibly good for $T=0.3$ and $0.25$ for all the different $t$'s; it is also good for $T=0.2$. The fitting for $T=0.15$ is good only for $t=1$ and $0.9$; it deteriorates for smaller $t$'s, and for $t=0.6$, it doesn't fit all. At $T=0.1$, we don't get a good fit to the LK formula for any of the $t$'s considered here. For $t\lesssim 0.6$, the oscillations nearly die off, and the fitting doesn't work. The general pattern noted here is that this LK fitting gets better with increasing $t$ and $T$. This empirical analysis suggests that the quantum oscillations of specific heat in Kondo insulators exhibit Lifshitz-Kosevish like thermal behaviour for smaller Kondo couplings and higher temperatures, and the frequencies of these oscillations come from the surface of charge gap. 

\begin{figure*}[htbp]
	\centering
	\includegraphics[width=\textwidth]{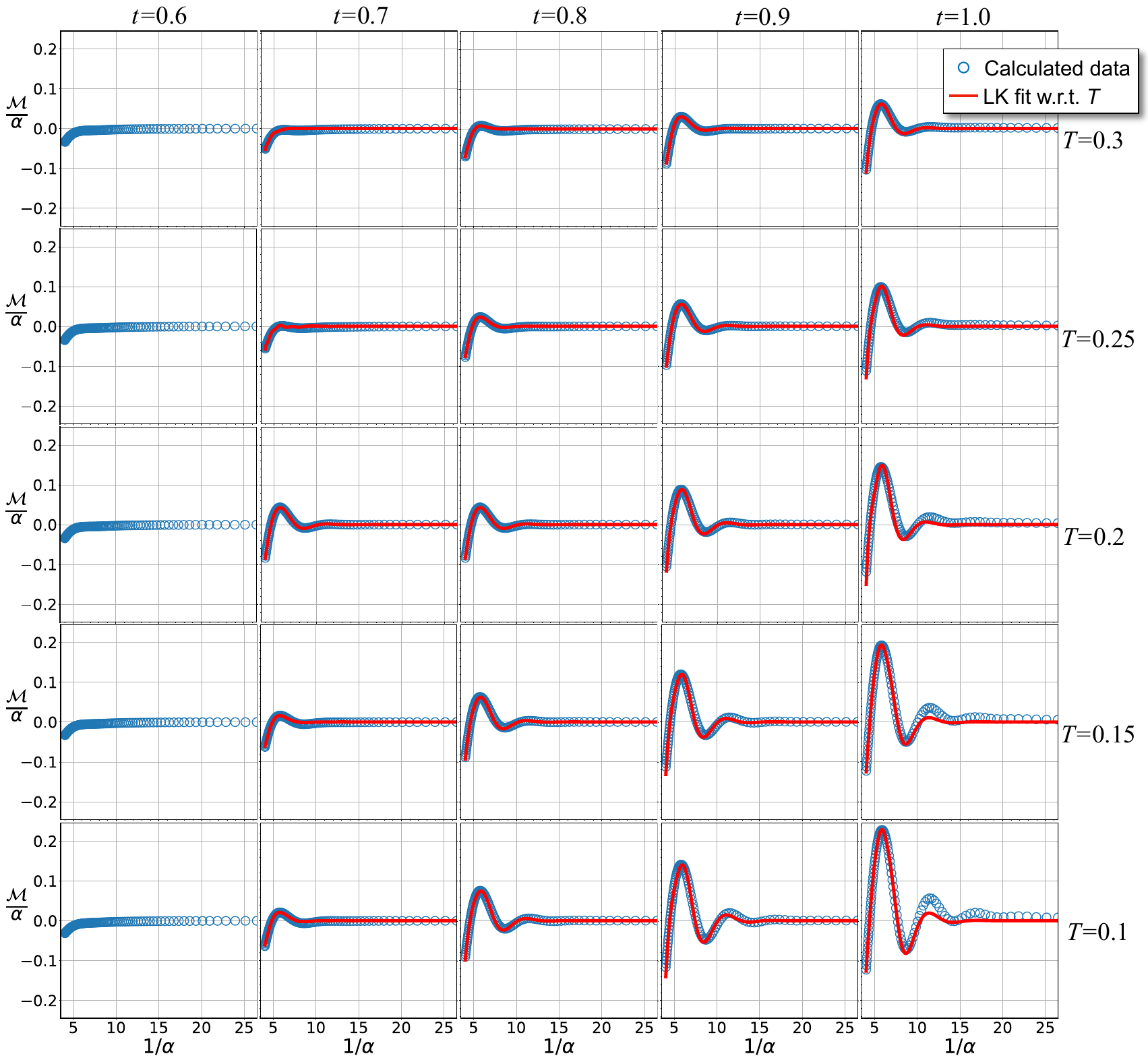}
	\caption{Magnetic quantum oscillations fitted for their thermal behaviour to the Lifshitz-Kosevich formula, Eq.~\eqref{eq:LK-M}, with two frequencies, $A_0$ and $A_\pi$, coming from the surface of charge gap.}
	\label{fig:LK_T_mag}
\end{figure*}

In Ref.~\cite{pushkar2023low}, we had similarly found that the magnetic quantum oscillations too exhibit Lifshitz-Kosevich like behaviour. But, there, we had used the value of 0.185 for the frequency. So, now, we take a relook at the magnetic oscillations by fitting them to the LK formula, Eq.~\eqref{eq:LK-M}, with two frequencies given by the surface of charge gap. We find that this two-frequency LK fit describes the magnetic quantum oscillations very well in most cases, as presented in Fig.~\ref{fig:LK_T_mag}, with an improvement over the previous single frequency fitting. 

\begin{figure*}[htbp]
	\centering
	\includegraphics[width=\textwidth]{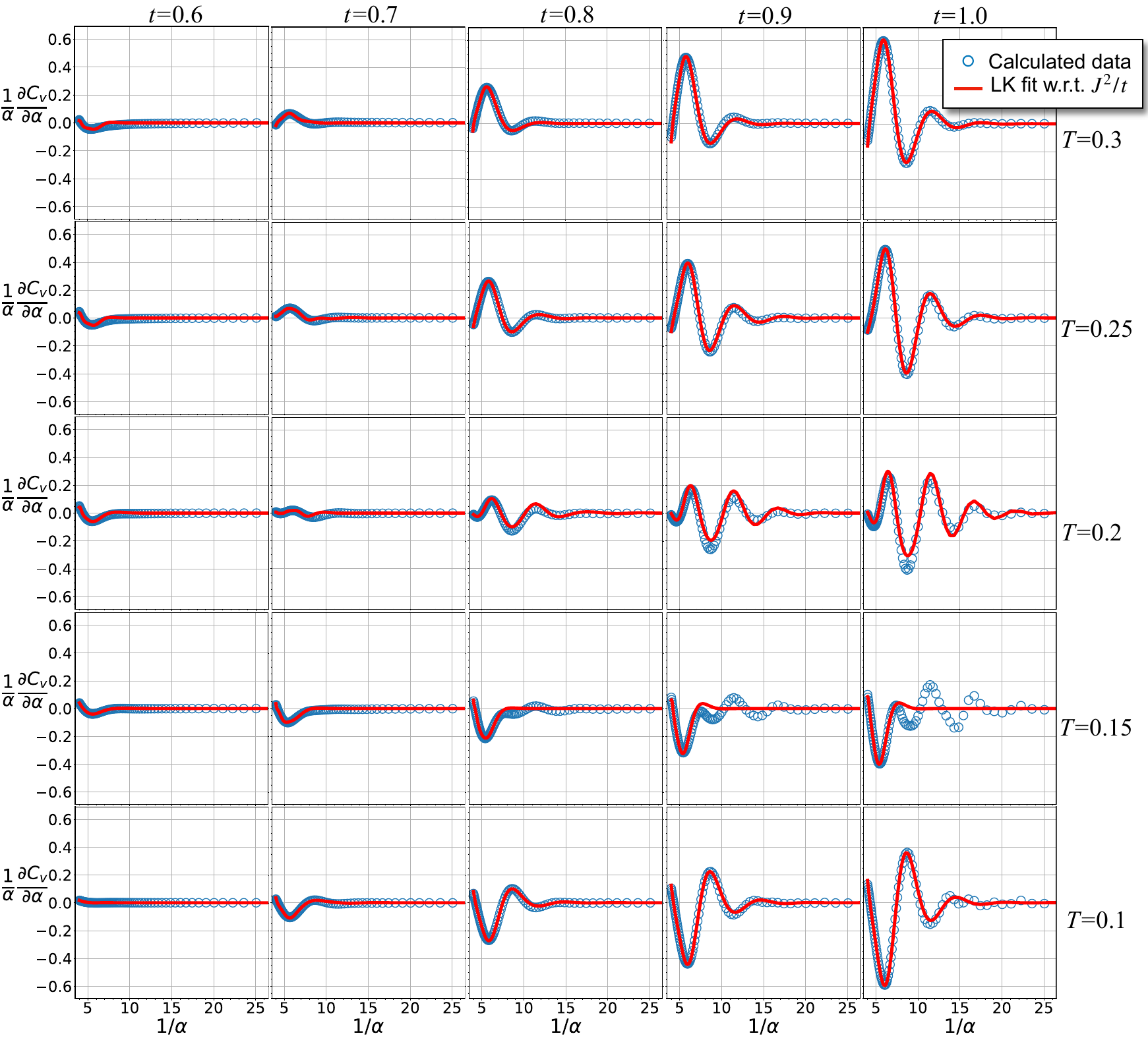}
	\caption{Specific heat oscillations fitted empirically to Eq.~\eqref{eq:LK-cv} with $T$ replaced by $J^2/t$ for the two frequencies given in Table~\ref{tab:Area}. This is to understand the behaviour these quantum oscillations with respect to the quantum fluctuations characterized by $J^2/t$. Clearly, the fitting is quite agreeable in most cases.}  
	\label{fig:LK_tinv_Cv}
\end{figure*}

\begin{figure*}[htbp]
	\centering
	\includegraphics[width=\textwidth]{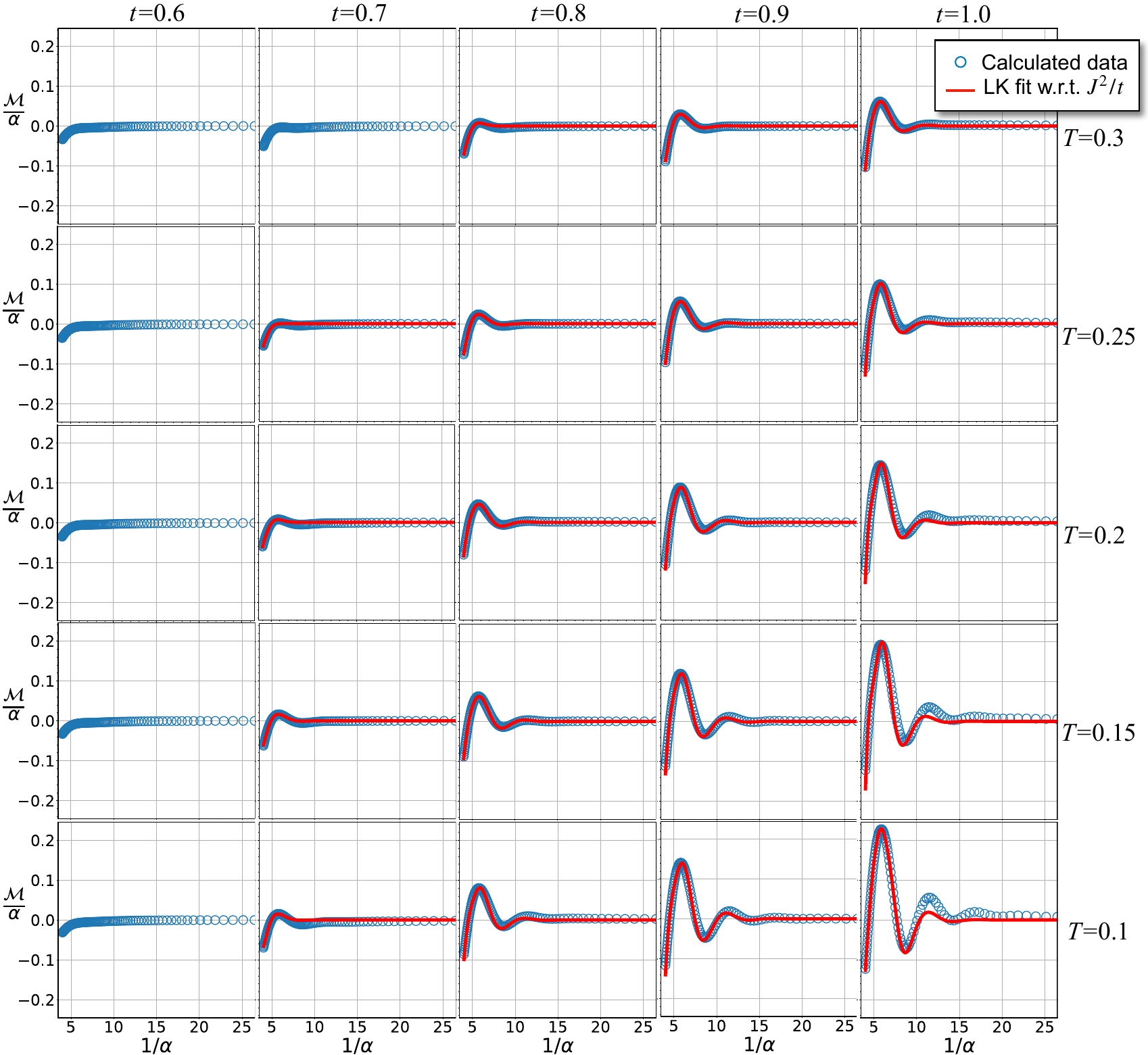}
	\caption{Magnetic quantum oscillations fitted to Eq.~\eqref{eq:LK-M} with $T$ replaced by $J^2/t$ for the two frequencies listed in Table~\ref{tab:Area}. Clearly, these oscillations follow Lifshitz-Kosevich like behaviour also with respect to the quantum fluctuations.}
	\label{fig:LK_tinv_mag}
\end{figure*}

In Ref.~\cite{pushkar2023low}, we had also found that an empirical Lifshitz-Kosevich formula in which $T$ is replaced by $J^2/t$ (a measure of quantum fluctuations) describes the evolution of magnetic quantum oscillations with hopping quite well. Here we check that with a two-frequency fit of the specific heat and magnetic oscillations to the modified versions of Eqs.~\eqref{eq:LK-M} and~\eqref{eq:LK-cv} with $T$ replaced by $J^2/t$. The results, presented in Fig.~\ref{fig:LK_tinv_Cv} for specific heat and in Fig.~\ref{fig:LK_tinv_mag} for magnetization, are indeed very good. 

The overall effectiveness of the two-frequency Lifshitz-Kosevch fits presented in Figs.~\ref{fig:LK_T_Cv} to~\ref{fig:LK_tinv_mag} provides an empirical support to the basic idea of this theory that the frequencies of quantum oscillations in a Kondo insulator come from the surface of charge gap, a memory of the Fermi surface in its metallic state. From this analysis, we also expect the quantum oscillations in Kondo insulators to exhibit Lifshitz-Kosevich like behaviour with respect to thermal as well as quantum fluctuations, at higher temperatures and for weaker Kondo couplings. This is in a way consistent with the reported thermal behaviour of quantum oscillations in Kondo insulators; for instance, the magnetic oscillations in \ce{SmB6} are reported to show deviations from the LK behaviour only below about 1 K~\cite{tan2015unconventional,hartstein2018fermi} and in \ce{YbB12} below 60 mK~\cite{Liu_2018,Xiang_2018}.

\section{\label{sec:sum}Summary}
In this paper, we investigated the quantum oscillations of specific heat in Kondo insulators by means of a theory formulated in Refs.~\cite{Ram2017,pushkar2023low}. As for magnetization, the specific heat too is found to exhibit quantum oscillations in the insulating phases with inverted quasiparticle dispersion; in fact, the specific heat oscillations turned out to be even more pronounced than the magnetic ones over a wider range of Kondo couplings and temperatures. It lends further support to the basic finding of this theory that the inversion is a key determinant for the quantum oscillations to occur in Kondo insulators. Interestingly, we realized that the field derivative of specific heat better represents the quantum oscillations of specific heat, which can be directly compared with magnetic quantum oscillations. Akin to magnetization, the quantum oscillations of specific heat are also found to exhibit Lifshitz-Kosevich like behaviour with respect to temperature as well as the parameter $J^2/t$ characterizing quantum fluctuations. We showed this empirically by fitting the calculated quantum oscillations to the Lifshitz-Kosevich formulae, Eq.~\eqref{eq:LK-M} and~\eqref{eq:LK-cv}, with two frequencies corresponding to the extremal areas on the surface of charge gap. It also shows that the frequencies of quantum oscillations in Kondo insulator come from the surface of charge gap. 

\section{Acknowledgements}
We acknowledge the DST-FIST funded HPC cluster at the School of Physical Sciences, JNU for computations.

\appendix*
\section{Diagonalization of $\Hcal_{c,\alpha}$}
Here we describe the procedure for computing the field dependent internal energy, $u_\alpha = \langle \Hhat_{c,\alpha}\rangle/L$, from the minimal effective model, Eq.~\eqref{eq:Hcalpha}, for the charge dynamics of the half-filled KLM in  magnetic field.

The simple cubic lattice, with distinct $A$ and $B$ sublattices, is described here by the primitive vectors $\vec{a}_1=2a\hat{x}$, $\vec{a}_2=a(-\hat{x}+\hat{y})$ and $\vec{a}_3=a(-\hat{x}+\hat{z})$, with a two site basis made of an $A$ site and its $B$ neighbour at $a\hat{x}$.  The corresponding Bravais lattice of $A$ sites is given by the position vectors $\vec{r}=m\vec{a}_1+n\vec{a}_2+l\vec{a}_3 \equiv (m,n,l)$. For a given $A$ site at $(m,n,l)$, its nearest $B$ neighbours at $\pm a\hat{x}$ are $(m,n,l)$ and $(m-1,n,l)$, at $\pm a\hat{y}$ are $(m,n+1,l)$ and $(m-1,n-1,l)$, and those at $\pm a\hat{z}$ are $(m,n,l+1)$ and $(m-1,n,l-1)$. In this notation, Eq.~\eqref{eq:Hcalpha} in terms of the spinless fermion operators reads as:
\begin{widetext}
\begin{align}
	\Hcal_{c,\alpha} = &-\frac{t}{2} \sum_{m=1}^{L_1}\sum_{n=1}^{L_2}\sum_{l=1}^{L_3}
	\bigg[\cos(2\pi\alpha n)\left\{(1+\rho_1)[(\ahat^\dagger_{m,n,l}\bhat_{m,n,l}+\bhat^\dagger_{m,n,l}\ahat_{m,n,l})+(\ahat^\dagger_{m,n,l}\bhat_{m-1,n,l}+\bhat^\dagger_{m-1,n,l}\ahat_{m,n,l,})]\right\}\nonumber\\
	&+\cos(2\pi\alpha n)\left\{(1-\rho_1)[(\ahat^\dagger_{m,n,l}\bhat^\dagger_{m,n,l}+\bhat_{m,n,l}\ahat_{m,n,l})+(\ahat^\dagger_{m,n,l}\bhat^\dagger_{m-1,n,l}+\bhat_{m-1,n,l}\ahat_{m,n,l,})]\right\}\nonumber\\
	&+(1+\rho_1)\left[(\ahat^\dagger_{m,n,l}\bhat_{m,n+1,l}+\bhat^\dagger_{m,n+1,l}\ahat_{m,n,l})+(\ahat^\dagger_{m,n,l}\bhat_{m-1,n-1,l}+\bhat^\dagger_{m-1,n-1,l}\ahat_{m,n,l,})\right]\nonumber\\
	&+(1+\rho_1)\left[(\ahat^\dagger_{m,n,l}\bhat_{m,n,l+1}+\bhat^\dagger_{m,n,l+1}\ahat_{m,n,l})+(\ahat^\dagger_{m,n,l}\bhat_{m-1,n,l-1}+\bhat^\dagger_{m-1,n,l-1}\ahat_{m,n,l,})\right]\nonumber\\
	&+(1-\rho_1)\left[(\ahat^\dagger_{m,n,l}\bhat^\dagger_{m,n+1,l}+\bhat_{m,n+1,l}\ahat_{m,n,l})+(\ahat^\dagger_{m,n,l}\bhat^\dagger_{m-1,n-1,l}+\bhat_{m-1,n-1,l}\ahat_{m,n,l,})\right]\nonumber\\
	&+(1-\rho_1)\left[(\ahat^\dagger_{m,n,l}\bhat^\dagger_{m,n,l+1}+\bhat_{m,n,l+1}\ahat_{m,n,l})+(\ahat^\dagger_{m,n,l}\bhat^\dagger_{m-1,n,l-1}+\bhat_{m-1,n,l-1}\ahat_{m,n,l,})\right]\bigg]\nonumber\\
	&+\frac{J\rho_0}{4}\sum_{m=1}^{L_1}\sum_{n=1}^{L_2}\sum_{l=1}^{L_3}\left(\ahat^\dagger_{m,n,l}\ahat_{m,n,l}+\bhat^\dagger_{m,n,l}\bhat_{m,n,l}\right)
\end{align}
\end{widetext}
with $r_y=n$ in the cosines carrying the magnetic field as $\alpha$. Moreover, $2L_1L_2L_3=L$ is the total number of sites.

Doing Fourier transformation along the free directions $\vec{a}_1$ and $\vec{a}_3$  
\begin{subequations}
	\begin{align}
	\ahat_{m,n,l} & =\frac{1}{\sqrt{L_1L_3}}\sum_{k_1,k_3}e^{i(k_1 m + k_3 l)} \, \ahat_{k_1,n,k_3} \label{eq:FTa_m} \\
		\bhat_{m,n,l} & =\frac{1}{\sqrt{L_1L_3}}\sum_{k_1,k_3}e^{i[k_1(m+\frac{1}{2})+k_3l]} \, \bhat_{k_1,n,k_3} \label{eq:FTb_m}
	\end{align}
\end{subequations}
leads to
\begin{align}
	\Hcal_{c,\alpha}=&-\frac{t}{2}\sum_{k_1,k_3}\sum_{n=1}^{L_2}\bigg[2\Big\{\cos(2\pi\alpha n)\cos{\frac{k_1}{2}}+\cos\Big(\frac{k_1}{2}+k_3\Big)\Big\}\nonumber\\
	&\times\Big[(1+\rho_1)(\ahat^\dagger_{k_1,n,k_3}\bhat_{k_1,n,k_3}+h.c.)\nonumber\\
	&\hspace{.6cm}+(1-\rho_1)(\ahat^\dagger_{k_1,n,k_3}\bhat^\dagger_{-k_1,n,-k_3}+h.c.)\Big]\nonumber\\
	&+(1+\rho_1)\Big[(e^{i\frac{k_1}{2}}\ahat^\dagger_{k_1,n,k_3}\bhat_{k_1,n+1,k_3}+h.c.)\nonumber\\
	&\hspace{1.8cm}+(e^{-i\frac{k_1}{2}}\ahat^\dagger_{k_1,n,k_3}\bhat_{k_1,n-1,k_3}+h.c.)\Big]\nonumber\\
	&+(1-\rho_1)\Big[(e^{i\frac{k_1}{2}}\ahat^\dagger_{k_1,n,k_3}\bhat^\dagger_{-k_1,n+1,-k_3}+h.c.)\nonumber\\&\hspace{1.8cm}+(e^{-i\frac{k_1}{2}}\ahat^\dagger_{k_1,n,k_3}\bhat^\dagger_{-k_1,n-1,-k_3}+h.c.)\Big]\bigg]\nonumber\\
	&+\frac{J\rho_0}{4}\sum_{k_1,k_3}\sum_{n=1}^{L_2}\left(\ahat^\dagger_{k_1,n,k_3}\ahat_{k_1,n,k_3}+\bhat^\dagger_{k_1,n,k_3}\bhat_{k_1,n,k_3}\right).
	\label{eq:Hc_k3}
\end{align}

Note that $\Hcal_{c,\alpha}$ is not periodic along $n$ in general. But for $\alpha=p/q$, where $p=1,2,\dots, q$ with $q$ as a prime number, it becomes periodic along $n$ with period $q$. For this choice of $\alpha$, each super-unit-cell labelled by integer $n^\prime$ consists of $q$ adjacent sites indexed as $n^{\prime\prime}=1,2,3,\dots,q$. Hence, $n=q(n^\prime-1)+n^{\prime\prime}$ with $n^\prime=1,2,3,\dots,L_2^\prime$ such that $L_2=q L^\prime_2$ and 
$\sum_{n=1}^{L_2}=\sum_{n^\prime=1}^{L_2^\prime}\sum_{n^{\prime\prime}=1}^{q}$. Now, we can do the Fourier transform with respect to $n^\prime$ as
\begin{subequations}
	\begin{align}
		\ahat_{k_1,n,k_3} &=\frac{1}{\sqrt{L^\prime_2}}\sum_{k_2}e^{i[q(n'-1)+n^{\prime\prime}]k_2}\ahat_{\vec{k},n^{\prime\prime}} \label{eq:FTa_n} \\
		\bhat_{k_1,n,k_3} &=\frac{1}{\sqrt{L^\prime_2}}\sum_{k_2}e^{i[q(n'-1)+n^{\prime\prime}]k_2}\bhat_{\vec{k},n^{\prime\prime}}
		\label{eq:FTb_n}
	\end{align}
\end{subequations}
where $\vec{k} = (k_1,k_2,k_3)$. With this, $\Hcal_{c,\alpha}$ becomes 
\begin{align}
	\Hcal_{c,\alpha}=&\sum_{\k}\sum_{n^{\prime\prime}=1}^{q}\Big[\Lambda_{\vec{k},n^{\prime\prime}}(1+\rho_1)[\ahat^\dagger_{\k,n^{\prime\prime}}~\bhat_{\k,n^{\prime\prime}}+h.c.]\nonumber\\
	&+\Lambda_{\vec{k},n^{\prime\prime}}(1-\rho_1)[\ahat^\dagger_{\k,n^{\prime\prime}}~\bhat^\dagger_{-\k,n^{\prime\prime}}+h.c.]\nonumber\\
	&+(1+\rho_1)[\lambda_k\ahat^\dagger_{\k,n^{\prime\prime}}~\bhat_{\k,n^{\prime\prime}+1}+ 
	\lambda^*_k\ahat^\dagger_{\k,n^{\prime\prime}}~\bhat_{\k,n^{\prime\prime}-1}+h.c.]\nonumber\\
	&+(1-\rho_1)[\lambda_k\ahat^\dagger_{\k,n^{\prime\prime}}~\bhat^\dagger_{-\k,n^{\prime\prime}+1} + 
	\lambda^*_k\ahat^\dagger_{\k,n^{\prime\prime}}~\bhat^\dagger_{-\k,n^{\prime\prime}-1}+h.c.]\Big]\nonumber\\
	&+\frac{J\rho_0}{4}\sum_{\k}\sum_{n_q=1}^{q}\left(\ahat^\dagger_{\k,n^{\prime\prime}}~\ahat_{\k,n^{\prime\prime}}+\bhat^\dagger_{\k,n^{\prime\prime}}~\bhat_{\k,n^{\prime\prime}}\right) 
	\label{eq:Hc_k2}
\end{align}
where
\begin{align}
	&\Lambda_{\vec{k},n^{\prime\prime}}=-t[\cos(2\pi\alpha n^{\prime\prime})\cos(k_1/2)+\cos(k_1/2+k_3)]\nonumber
\end{align}
and
\begin{align}
	&\lambda_k=-\frac{t}{2}e^{i(k_1/2+k_2)}.\nonumber
\end{align}
The $\Hcal_{c,\alpha}$ can now be written in the Nambu basis as
\begin{equation}
	\Hcal_{c,\alpha}=\frac{J\rho_0}{8}L+\frac{1}{2}\sum_{\k}\Psi^\dagger_{\k}\Hcal_{\k}\Psi_{\k}
	\label{eq:Hcalpa_nambu}
\end{equation}
where $\Psi^\dagger_{\k}=[\ahat^\dagger_{\k,1},\ahat^\dagger_{\k,2},\dots,\ahat^\dagger_{\k,q},\bhat^\dagger_{\k,1},\bhat^\dagger_{\k,2},\dots,\bhat^\dagger_{\k,q},\ahat_{-\k,1},\\ \ahat_{-\k,2},\dots,\ahat_{-\k,q},\bhat_{-\k,1},\bhat_{-\k,2},\dots,\bhat_{-\k,q}]$
and $\Hcal_{\k}$ is the $4q\times4q$ matrix given below.

\begin{equation}
	\Hcal_{k}=
	\begin{pmatrix}
		A & B_{\k} & 0 & \tilde{B}_{\k}\\
		B_{\k} & A & -\tilde{B}^*_{\k} & 0\\
		0 & -\tilde{B}^*_{\k} & -A & -B^*_{\k}\\
		\tilde{B}_{\k} & 0 & -{B}^*_{\k} & -A
	\end{pmatrix}
	\label{eq:Hk_mat}
\end{equation}
Here $A=\frac{J\rho_0}{8}\mathbb{I}_{q\times q}$ with $\mathbb{I}_{q\times q}$ as the $q \times q$ identity matrix, 
\begin{widetext}	
\begin{align}
		B=
		\begin{pmatrix}
			\Lambda_k(1)(1+\rho_1) & \lambda_k(1+\rho_1) & \cdots & \cdots & \cdots & \lambda^*_k(1+\rho)\\
			\lambda^*_k(1+\rho_1) & \Lambda_k(2)(1+\rho_1) & \lambda_k(1+\rho_1) & \cdots & \cdots & \vdots \\
			\vdots & \lambda^*_k(1+\rho_1) & \Lambda_k(3)(1+\rho_1) & \lambda_k(1+\rho_1) & \cdots & \vdots \\
			\vdots & \ddots & \ddots & \ddots\ & \cdots & \vdots\\
			\vdots & \cdots & \ddots & \ddots & \ddots & \vdots \\
			\vdots & \cdots & \cdots & \lambda^*_k(1+\rho_1) & \Lambda_k(q-1)(1+\rho_1) & \lambda_k(1+\rho_1)\\
			\lambda_k(1+\rho_1) & \cdots & \cdots & \cdots & \lambda^*_k(1+\rho_1) & \Lambda_k(q)(1+\rho_1)
		\end{pmatrix}
\end{align}

and

\begin{align}
	\tilde{B}=
	\begin{pmatrix}
		\Lambda_k(1)(1-\rho_1) & \lambda_k(1-\rho_1) & \cdots & \cdots & \cdots & \lambda^*_k(1-\rho)\\
		\lambda^*_k(1-\rho_1) & \Lambda_k(2)(1-\rho_1) & \lambda_k(1-\rho_1) & \cdots & \cdots & \vdots \\
		\vdots & \lambda^*_k(1-\rho_1) & \Lambda_k(3)(1-\rho_1) & \lambda_k(1-\rho_1) & \cdots & \vdots \\
		\vdots & \ddots & \ddots & \ddots\ & \cdots & \vdots\\
		\vdots & \cdots & \ddots & \ddots & \ddots & \vdots \\
		\vdots & \cdots & \cdots & \lambda^*_k(1-\rho_1) & \Lambda_k(q-1)(1-\rho_1) & \lambda_k(1-\rho_1)\\
		\lambda_k(1-\rho_1) & \cdots & \cdots & \cdots & \lambda^*_k(1-\rho_1) & \Lambda_k(q)(1-\rho_1)
	\end{pmatrix}
\end{align}
\end{widetext}
We can see that $\tilde{B}^*$ and $B^*$ are simply the transpose of $\tilde{B}$ and $B$ matrices.

The $\Hcal_{c,\alpha}$ can be diagonalized numerically by doing Bogoliubov transformation to a new Nambu basis: $\tilde{\Psi}^\dagger_{\k}=[{\eta}^\dagger_{\k,1},{\eta}^\dagger_{\k,2},\dots,{\eta}^\dagger_{\k,2q},~{\eta}_{-\k,1},{\eta}_{-\k,2},\dots,\dots,{\eta}_{-\k,2q}]$, such that 
\begin{equation}
	\Hcal_{c,\alpha}=\frac{J\rho_0}{8}L+\sum_{\k}\sum_{\nu=1}^{2q}\epsilon_{\vec{k},\nu}\left({\eta}^\dagger_{\k,{\nu}}{\eta}_{\k,{\nu}}-\frac{1}{2}\right)
	\label{eq:Hcalpha_fin}
\end{equation}
where $\epsilon_{\k,1},~\dots~, \epsilon_{\k,2q}$ are the positive eigenvalues of $\mathcal{H}_{\k}$, Eq.~\eqref{eq:Hk_mat}, in the increasing order. The internal energy per site, therefore is
\begin{equation}
	{u}_\alpha=\frac{\langle \Hcal_{c,\alpha}\rangle}{L}=\frac{J\rho_0}{8}-\frac{1}{2L}\sum_{\vec{k}}\sum_{\nu=1}^{2q}\epsilon_{\vec{k},\nu}\tanh \left(\frac{\beta\epsilon_{\vec{k},\nu}}{2}\right)
	\label{eq:Ucalpha_fin}
\end{equation}
Similarly, the free energy can be written as,
\begin{align}
	\mathcal{F}_\alpha&=-\frac{1}{\beta}\log \{\tr(e^{-\beta\Hcal_{c,\alpha}})\}\nonumber\\
	&=\frac{J\rho_0 L}{8}-\sum_{\vec{k}}\sum_{\nu=1}^{2q}\frac{\epsilon_{\vec{k},\nu}}{2}-\frac{1}{\beta}\sum_{\vec{k}}\sum_{\nu=1}^{2q}\log (1+e^{-\beta \epsilon_{\vec{k},\nu}})
	\label{eq:Fcalpha}
\end{align}
Using, \eqref{eq:Ucalpha_fin} and \eqref{eq:Fcalpha}, we can calculate the specific-heat and magnetization for every $\alpha$ respectively.

\bibliography{KLM_Bibliography.bib}

\end{document}